\documentclass[twocolumn,showpacs,pre]{revtex4}
\usepackage{latexsym}
\begin{document}
\title{The effect of shear on persistence in coarsening systems}
\author{N. P. Rapapa}
\author{A. J. Bray}
\affiliation{Department of Physics and Astronomy, The University, 
Oxford Road, Manchester M13 9PL, UK}
\date{\today}

\begin{abstract}
We  analytically study  the  effect of  a  uniform shear  flow on  the
persistence properties  of coarsening  systems.  The study  is carried
out  within the  anisotropic Ohta-Jasnow-Kawasaki  (OJK) approximation
for a system  with nonconserved scalar order parameter.   We find that
the persistence  exponent $\theta$ has  a non-trivial value: $\theta =
0.5034\ldots$ in  space dimension  $d=3$, and $\theta  = 0.2406\ldots$
for  $d=2$, the latter  being exactly  twice the  value found  for the
unsheared  system in  $d=1$.  We also  find  that the  autocorrelation
exponent $\lambda$ is affected by shear in $d=3$ but not in $d=2$.
\end{abstract}


\maketitle

\section{Introduction}
The phenomenon of persistence  in nonequilibrium systems has attracted
considerable   interest   in   recent  years   \cite{Majumdar},   both
theoretically \cite{Rud,Sire,Ehrhardt}      and        experimentally
\cite{Marcos,Yurke,Tam,Wong,Dougherty}.   The persistence probability,
$P_0(t)$, of a fluctuating, spatially homogeneous nonequilibrium field
is the  probability that the field  $X(t)$ at a given  space point has
not changed sign up to time $t$.  This probability typically decays as
a  power  law,  $P_0(t)\sim  t^{-\theta}$  at late  times,  where  the
persistence  exponent  $\theta$ has  in  general  a nontrivial  value.
Persistence has been studied in  a considerable number of systems such
as  simple diffusion  from random  initial  conditions, phase-ordering
kinetics,  fluctuating  interfaces  and  reaction-diffusion  processes
\cite{Majumdar}.

Experiments to  determine persistence exponents have  been carried out
in  the  context  of  breath figures  \cite{Marcos},  liquid  crystals
\cite{Yurke},  soap froths  \cite{Tam},  diffusion of  Xe  gas in  one
dimension   \cite{Wong}   and  fluctuating   monatomic   steps  on   a
metal/semiconductor adsorption  system Si-Al surface \cite{Dougherty}.
Many  of these  cases are  examples of  coarsening phenomena,  where a
characteristic length scale increases  with time as the system relaxes
towards an equililibrium  that it attains only after  infinite time in
the  thermodynamic limit.  The experimental  results are  generally in
good  quantitative agreement with  (exact or  approximate) theoretical
predictions.

A classic example of a  coarsening phenomenon is the dynamics of phase
ordering,   where   a   system   is   quenched   from   a   disordered
high-temperature phase  into an ordered low-temperature  phase. In the
simplest case  of a two-phase  system, domains of the  two equilibrium
phases form and  grow with time. The characteristic  length scale at a
given  time is  the typical  scale of  the domain  structure  that has
formed at that time.  The coarsening dynamics is usually characterised
by  a   form  of  dynamical   scaling,  in  which  the   system  looks
statistically similar at different  times apart from an overall change
of scale \cite{Bray}.

Recently there has  been interest in the effect of  shear in a variety
of systems  such as  macromolecules, binary fluids  and self-assembled
fluids  \cite{Onuki}.  Shear introduces  anisotropy  into the  spatial
structure. For  systems undergoing phase  ordering in the  presence of
shear,  the domain  growth  becomes anisotropic  and  this results  in
different growth  exponents for the structure  along and perpendicular
to  the flow. At  present it  is not  clear whether  shear leads  to a
stationary   steady   state,  or   whether   domain  growth   proceeds
indefinitely  at asymptotically  large times  \cite{Cates}.  Shear may
also induce phase transitions: for example, shear-induced shift of the
phase transition  temperature in the microphase  separation of diblock
copolymers has been observed \cite{Milner}.

In this paper  we analytically study the effect  of an imposed uniform
shear  flow on  persistence for  the simplest  case of  a nonconserved
scalar   order    parameter.    We   exploit   a    version   of   the
Ohta-Jasnow-Kawasaki (OJK) approximation in  phase-ordering   kinetics
\cite{Bray},  modified to account  for the  anisotropy induced  by the
shear \cite{Cavagna}.  Persistence is  defined here as the probability
that a  point {\em comoving  with the flow}  has remained in  the same
phase up  to time $t$.  We  employ an approach  called the independent
interval  approximation  (IIA) which  has  been  successfully used  to
obtain rather  accurate values for persistence  exponents in unsheared
systems  \cite{Majumdar}.  This procedure  assumes that  the intervals
between zeros of the process $X(t)$ are statistically independent when
measured in  the mapped time  variable $T =  \ln t$. We find  that the
persistence  exponent   $\theta$  is  nontrivial   and  dimensionality
dependent.   For $d=3$  we find  $\theta \simeq  0.5034$,  compared to
$\simeq 0.2358$ in the unsheared case\cite{Sire}, while $\theta \simeq
0.2406$  for $d=2$ compared  to $\theta  \simeq 0.1862$  without shear
\cite{Sire}.  Remarkably, the value  of $\theta$  in $d=2$  is exactly
twice the value obtained for the unsheared system in $d=1$ \cite{Sire}
using similar  methods. There is  a technical subtlety in  $d=2$ which
requires a careful definition of the persistence probability.  In both
$d=2$  and  $d=3$ the  shear  increases  the  persistence exponent.

The paper is organised as follows.  In the next section the OJK theory
is introduced  and the autocorrelation function, which  is a necessary
input  to  the IIA  calculation,  is  obtained  for $d=3$  and  $d=2$.
Section III contains a brief outline  of the IIA, the results of which
for  the  sheared problem  are  presented  in  section IV.  Concluding
remarks are given in section IV.

\section{The OJK Theory}
We consider a nonconserved scalar order parameter $\phi(\vec{x},t)$ 
evolving via the time-dependent Ginzburg-Landau equation \cite{Bray}
\begin{equation}
\frac{\partial \phi(\vec{x},t)}{\partial t}  
= \nabla^2 \phi(\vec{x},t) - V'(\phi), 
\label{TDGL}
\end{equation}
where $V(\phi)$  is a symmetric double-well  potential. The assumption
that the  thickness $\xi$ of  the interface separating the  domains is
much  smaller than  the size  of the  domains allows  one to  write an
equation  of   motion  for   the  interface,  called   the  Allen-Cahn
equation \cite{Allen}.   The  velocity   $v$  of   the   interface  is
proportional to the local curvature and given by
\begin{equation}
v(\vec{x},t) = -\vec{\nabla}.\vec{n}(\vec{x},t), 
\label{vel}
\end{equation}
where $\vec{n}(\vec{x},t)$ is the unit vector normal to the interface, 
defined in the direction of increasing order parameter. The normal vector 
can be written in general as
\begin{equation}
\vec{n}(\vec{x},t)=\frac{\vec{\nabla}m(\vec{x},t)}{\mid \vec{\nabla}
m(\vec{x},t)\mid} 
\label{norm}
\end{equation}
where $m(\vec{x},t)$ is the smooth field that has the same sign as the
order parameter $\phi$ and vanishes at the interfaces (where the order
parameter vanishes). It  is easier to work with  an equation of motion
for $m(\vec{x},t)$ than for $\phi(\vec{x},t)$, an idea that is exploited
in the OJK theory \cite{Ohta}.

By considering  a frame  locally comoving with  the interface,  with a
space-uniform shear in the $y$-direction and flow in the $x$-direction
(i.e.\  the  fluid  velocity   profile  is  given  by  $\vec{u}=\gamma
y\vec{e}_x$),  where   $\gamma$  is   the  constant  shear   rate  and
$\vec{e}_x$ is the unit vector in the flow direction, the OJK equation
for the field $m(\vec{x},t)$ becomes \cite{Cavagna}
\begin{eqnarray}
\frac{\partial m(\vec{x},t)}{\partial  t} &+& \gamma  y \frac{\partial
m(\vec{x},t)}{\partial  x} =  \nonumber \\  \nabla^2  m(\vec{x},t) &-&
\sum_{a,b=1}^{d}D_{ab}(t)  \frac{\partial^2 m(\vec{x},t)}{\partial x_a
\partial x_b},
\label{phi}
\end{eqnarray}
where
\begin{equation}
D_{ab}(t)= \langle n_a n_b\rangle,
\label{phim}
\end{equation}
and $\langle ...\rangle$ denotes  average over initial conditions (or,
equivalently, over  space). The correct  equation for $m$  involves an
unaveraged  $D_{ab}$,   but  the   equation  is  then   nonlinear  and
intractable.  The essence of  the OJK approximation is the replacement
of the product $n_a n_b$ by  its average. For an isotropic system this
gives, by symmetry, $D_{ab} = \delta_{ab}/d$, and the equation for $m$
reduces to the diffusion equation. For the anisotropic sheared system,
however,   $D_{ab}(t)$   has   to  be   determined   self-consistently
\cite{Cavagna}. From Eq.\ (\ref{phim}), it follows that
\begin{equation}
\sum_{a=1}^{d} D_{aa}(t)=1.
\label{sumr}
\end{equation}
In $k$-space, Eq.\ (\ref{phi}) can be written as
\begin{eqnarray}
&&\frac{\partial m(\vec{k},t)}{\partial t} - \gamma k_x \frac{\partial
m(\vec{k},t)}{\partial  k_y}  =  \nonumber \\  &&\left(-\sum_{a=1}^{d}
{k_a}^2 + \sum_{a,b=1}^{d}D_{ab}(t)k_a k_b \right)m(\vec{k},t). 
\label{mksp}
\end{eqnarray}

\subsection{The case $d=3$}
We  now consider  the above  equation in  dimension $d=3$ and solve it  
via the following change of variables \cite{Cavagna},
\begin{equation}
(k_x, k_y, k_z, t)\rightarrow (q_x,q_y-\gamma k_x\tau,q_z,\tau),
\label{transf}
\end{equation}
with the introduction of an equivalent field
\begin{equation}
\mu (\vec{q},\tau)= m(\vec{k},t).
\end{equation}
The   left   hand  side   of   (\ref{mksp})   now  becomes   $\partial
\mu/\partial\tau$  and  as  a  result  equation  (\ref{mksp})  can  be
integrated  directly  to give  (after  transforming  back to  original
variables)
\begin{eqnarray}
m(\vec{k},t)=&&m(k_x,   k_y+\gamma  k_x  t,   k_z,  0)   \nonumber  \\
&&\times\exp\left[-\frac{1}{4}\sum_{ab=1}^{3}k_a M_{ab}(t) k_b\right],
\label{transwf}
\end{eqnarray}
with non-vanishing matrix elements 
\begin{eqnarray}
M_{11}(t)   &=&   R_{11}(t)+2\gamma   tR_{12}(t)+\gamma^2   t^2R_{22},
\nonumber  \\ M_{12}(t)  &=& R_{12}(t)+\gamma  t R_{22}(t),\nonumber\\
M_{22}(t) &=& R_{22}(t),\nonumber\\ M_{33}(t) &=& R_{33}(t),
\end{eqnarray}
where
\begin{eqnarray}
R_{11}(t)    &=&    4\int_0^{t}   dt'\left\{[1-D_{11}(t')]+    2\gamma
t'D_{12}(t')\right. \nonumber  \\ &&+ \left.\gamma^2t'^2[1-D_{22}(t')]
\right\},      \nonumber     \\     R_{12}(t)      &=&     4\int_0^{t}
dt'\left\{-D_{12}(t')-   \gamma   t'[1-D_{22}(t')]\right\},\nonumber\\
R_{22}(t)  &=&  4\int_0^{t} dt'\left\{1-D_{22}(t')\right\},\nonumber\\
R_{33}(t) &=& 4\int_0^{t} dt'\left\{1-D_{33}(t')\right\}.
\end{eqnarray}
Due  to the  symmetry of  the original  OJK equation  (\ref{phi}), the
terms  $R_{13}$,  $R_{23}$,  $M_{13}$  and $M_{23}$  all  vanish.  The
assumption that the initial  condition, $m(\vec{k},0)$, has a Gaussian
distribution, appropriate to a quench from the high-temperature phase,
is used throughout the paper.

In order to  use the IIA to investigate  the persistence properties of
the coarsening  system, it is first  necessary \cite{Majumdar,Sire} to
compute  the autocorrelation  function  of the  rescaled field  $X(t)=
m(\vec{x},t)/\langle    [m(\vec{x},t)]^2\rangle^{1/2}$,    which    is
constructed to have unit variance, using the initial correlator
\begin{equation} 
\langle m(\vec{x},0)m(\vec{x'},0)\rangle = \Delta\delta^d(\vec{x}-\vec{x'}).
\end{equation}
The  quantity $\langle  [m(\vec{x},t)]^2\rangle^{1/2}$  can easily  be
evaluated to give
\begin{equation}
\left\langle                       [m(\vec{x},t)]^2\right\rangle^{1/2}=
\left[\frac{\Delta}{(2\pi)^{3/2}}\,
\frac{1}{\sqrt{{\rm{Det}}\,\,M(t)}}\right]^{1/2}.
\label{normf}
\end{equation}

Turning now to the two-time correlator of $X(t)$, we recall that we want to 
calculate this correlator not at a fixed point in space, but at a point 
that is advected with the shear flow. Due to the shear, the  field  at  
the space-time point  $(x+\gamma yt_1,y,z,t_1)$ at time $t_1$ will be at  
the space-time point $(x+\gamma yt_2,y,z,t_2)$ at time
$t_2$. The autocorrelation function $a(t_1,t_2) = \left \langle X(t_1)
X(t_2)\right \rangle$ is therefore given by
\begin{widetext}
\begin{eqnarray}
a(t_1,t_2)=\left[\frac{(2\pi)^3}{\Delta^2}\sqrt{{\rm Det}\,M(t_1)
{\rm Det}\,M(t_2)}\right]^{1/2}\,\left\langle m(x+\gamma yt_1,y,z,t_1)
m(x+\gamma yt_2,y,z,t_2)\right\rangle.
\label{auto}
\end{eqnarray}
The  next step  is to  evaluate the  term $\langle  \cdot  \cdot \cdot
\rangle$  in the  above equation.  We note  that average  over initial
conditions in $k$-space implies
\begin{equation} 
\langle m(k_x,  k_y+\gamma k_x t_1,  k_z, 0) m(k'_x,  k'_y+\gamma k'_x
t_2,k'_z,0)\rangle = (2\pi)^d \Delta \delta(k_x+k'_x)   \nonumber
\delta(k_y+\gamma k_x t_1+k'_y+\gamma k'_x t_2)\delta( k_z+k'_z).
\label{normff}
\end{equation}
\end{widetext}

Using Eqs.\ (\ref{transwf}) and(\ref{normff}) we can evaluate the term
\begin{eqnarray} 
\langle             m(1)             m(2)\rangle&=&             \Delta
\sum_{k}^{}\exp\left[-\frac{1}{2}\sum_{a,b   =1}^{3}k_aB_{ab}k_b\right]
\nonumber        \\        &=       &\frac{\Delta}{(2\pi)^{3/2}}\,
\frac{1}{\sqrt{{\rm{Det}}\,\,B(t_1,t_2)}},
\label{finsl}
\end{eqnarray}
where
\begin{eqnarray}
B_{11}(t)          &=&          \left[M_{11}(t_1)+M_{11}(t_2)+\gamma^2
M_{22}(t_2)\times\left(t_2-t_1\right)^2\right.       \nonumber      \\
&&\left.   -2\gamma\left(t_2-t_1\right)M_{12}(t_2)\right]/2,\nonumber\\
B_{12}(t)                 &=&                [M_{12}(t_1)+M_{12}(t_2)-
\gamma\left(t_2-t_1\right)M_{22}(t_2)]/2,\nonumber\\   B_{22}(t)   &=&
[M_{22}(t_1)+M_{22}(t_2)]/2,\nonumber\\          B_{33}(t)         &=&
[M_{33}(t_1)+M_{33}(t_2)]/2, \nonumber\\ B_{13}(t) &=& B_{23}(t)=0.
\label{deal}
\end{eqnarray}

The notation $(1)$ and $(2)$ in (\ref{finsl}) denotes space-time points
$(x+\gamma     yt_1,y,z,t_1)$     and    $(x+\gamma     yt_2,y,z,t_2)$
respectively.   The  problem   is now   reduced  to   evaluating  the
determinants  of   the  matrices  $M(t)$  and   $B(t_1,t_2)$,  as  the
autocorrelation function $a(t_1,t_2)$ can now be written as
\begin{equation}
a(t_1,t_2)=\frac{\left[{\rm{Det}}\,M(t_1)\, 
{\rm{Det}} \,M(t_2)\right]^{1/4}}{\sqrt{{\rm{Det}}\,B(t_1,t_2)}}.
\label{ff2af}
\end{equation}
The terms  $M_{ab}(t)$ cannot be computed explicitly  for general $t$; 
only in  the scaling limit (i.e. $t\rightarrow \infty$) can one make
progress. In  this limit  it can  be shown that  (to leading  order for 
large $t$) \cite{Cavagna}
\begin{eqnarray}
M_{11}(t)  &=& \frac{4}{15}\gamma^2  t^3, \nonumber  \\  M_{12}(t) &=&
\frac{2}{5}\gamma   t^2,\nonumber\\    M_{22}(t)   &=&   \frac{4}{5}t,
\nonumber\\ M_{33}(t) &=& \frac{16}{5}t.
\label{cf}
\end{eqnarray}

Using   equations   (\ref{cf}),  the   determinants   of  $M(t)$   and
$B(t_1,t_2)$ can now be evaluated leading to
\begin{eqnarray}
{\rm{Det}}\,\,M(t)   =  &&\frac{64}{375}\gamma^2  t^5,   \nonumber  \\
{\rm{Det}}\,B(t_1,t_2)=&&\frac{8\gamma^2\left(t_2+t_1\right)^5}{125}\,
\left[4\left\{\frac{1}{3}-\frac{2{t_2}^2
t_1}{(t_2+t_1)^3}\right\}\right.              \nonumber             \\
&&\left. -\left\{1-\frac{2{t_2}^2}{(t_2+t_1)^2}\right\}^2\right].
\label{cfr}
\end{eqnarray}
The autocorrelation function follows:
\begin{eqnarray}
a(t_1,t_2)&&=\frac{1}{2}\,\frac{W^{5/4}(t_1,t_2)}{\left[1-\frac{3}{4}W^2
(t_1,t_2)\right]^{1/2}},\nonumber        \\       &&=\frac{1}{2}\,
\frac{{\rm{sech}}^{5/2}(T/2)}{\left[1-\frac{3}{4}{\rm{sech}}^4
(T/2)\right]^{1/2}},
\label{acfr}
\end{eqnarray}
where $W(t_1,t_2)=4t_1t_2/(t_2+t_1)^2$, $T=T_1-T_2$  and the final form 
follows   after   introducing   the   new   time   variable   $T_i   =
\ln t_i$. After this change of the time variable, the autocorrelation 
function depends only on the time difference $T_1-T_2$.  Since the process 
$X(T)$ is also gaussian, the process $X(T)$ is a gaussian stationary process. 
This will be the case whenever the autocorrelation function of $X$ depends 
on $t_1$ and $t_2$ only through the ratio $t_1/t_2$, i.e.\ when it exhibits 
a scaling form. 

\subsection{The case $d=2$}
For $d=2$, we follow the same analysis as for $d=3$ but with the change of 
variables
\begin{equation}
(k_x, k_y, t)\rightarrow (q_x,q_y-\gamma k_x\tau,\tau).
\label{transftd}
\end{equation}
The solution of Eq.\ (\ref{mksp}) in the large-$t$ limit is given by the 
$d=2$ analogue of Eq.\ (\ref{transwf}): 
\begin{eqnarray}
m(\vec{k},t)=&&m(k_x, k_y+\gamma k_x t, 0)\nonumber \\
&&\times\exp\left[-\frac{1}{4}\sum_{a,b=1}^{2}k_a M_{ab}(t) k_b\right].
\label{soltd}
\end{eqnarray}
The matrix elements $M_{ab}(t)$ can be evaluated for large $t$ using 
asymptotic analysis along the lines outlined in ref.\ \cite{Cavagna}, 
with the result 
\begin{eqnarray}
M_{11}(t) &=& 4\gamma t^2\sqrt{\ln\gamma t} - \frac{3\gamma t^2}
{\sqrt{\ln{\gamma t}}} \nonumber  \\
M_{12}(t) &=& 4t\sqrt{\ln{\gamma t}} -\frac{2t}{\sqrt{\ln{\gamma t}}}
\nonumber\\
M_{22}(t) &=& \frac{4}{\gamma}\sqrt{\ln{\gamma t}},  
\label{fd}
\end{eqnarray}
where we have retained just the leading subdominant terms, of relative 
order $1/\ln(\gamma t)$. 

The subleading  terms in $M_{11}(t)$ and $M_{12}(t)$  are necessary as
there  are  cancellations  to  leading  terms in  the  determinant  of
$M(t)$, which is given by ${\rm Det}\,M(t)=4t^2$. Using Eq.\ 
(\ref{soltd})  the  following   averages  can be calculated:
\begin{eqnarray}
&&\left\langle [m(\vec{x},t)]^2\right\rangle^{1/2}=
\left[\frac{\Delta}{2\pi}\,\frac{1}{\sqrt{{\rm Det\,M(t)}}}\right]^{1/2} 
= \frac{1}{2t}\,\sqrt{\frac{\Delta}{2\pi}}, \nonumber  \\
&&\langle m(1) m(2)\rangle= 
\frac{\Delta}{(2\pi)}\, \frac{1}{\sqrt{{\rm{Det}}\,\,B(t_1,t_2)}}.
\label{ftdso}
\end{eqnarray}
where the matrix elements $B_{ab}$ are  given  by the expressions in  
Eq.\ (\ref{deal}) but  with the corresponding $M_{ab}(t)$  given their  
by their $d=2$   equivalents   in  Eq.\ (\ref{fd}). The autocorrelation 
function  $a(t_1,t_2)$ for $d=2$ can  now be evaluated using the set of 
equations (\ref{ftdso}) to give
\begin{equation}
a(t_1,t_2)= \left[\frac{4t_1t_2}{t_1^2\left(1+\sqrt{\frac{\ln \gamma t_2}
{\ln \gamma t_1}}\right) +t_2^2\left(1+\sqrt{\frac{\ln \gamma t_1}
{\ln \gamma t_2}}\right)}\right]^{1/2}.
\label{2dres}
\end{equation}
\smallskip

Note that $a(t_1,t_2)$ given by  that Eq.\ (\ref{2dres}) does not have
a scaling form, i.e.\ it is not simply a function of $t_1/t_2$, due to
the logarithms.  However it does  a scaling {\em regime}. In the limit
$t_1  \to  \infty$,  $t_2   \to  \infty$,  with  $t_1/t_2$  fixed  but
arbitrary,  the ratio  of logarithms  can  be replaced  by unity  and
$a(t_1,t_2)$ depends  only on $t_1/t_2$  in this regime. In terms of 
the new time variable $T=\ln(t_2/t_1)$, one obtains
\begin{equation}
a(t_1,t_2) = \sqrt{{\rm sech}(T)},
\label{2dfinal}
\end{equation}
where $T=T_1-T_2$,  i.e. the process $X(T)$ becomes  stationary in the
defined scaling  limit. We will  use Eq.\ (\ref{2dfinal})  rather than
Eq.\ (\ref{2dres})  to extract $\theta$  for $d=2$, but one  must note
the  special   limit  taken   to  derive  (\ref{2dfinal})   where  the
persistence probbaility  $P(t_1,t_2)$ is the probability  that a point
moving with the flow has stayed  in the same phase between times $t_1$ 
and $t_2$. 

\section{The Independent Interval Approximation}

The  above analysis  in  both $d=3$  and  $d=2$ shows  that $X(t)$  is
stationary in the  new time variable $T$ (with  the caveat noted above
for  $d=2$).  We  note that  the  expected form  for the  probability,
$P_0(t)$, of  $X(t)$ having no  zeros between $t_1$ and  $t_2$, namely
$P_0\sim(t_1/t_2)^\theta$ for $t_2 \gg t_1$, becomes exponential decay
$P_0\sim e^{-\theta(T_2-T_1)}$ in the new time variable.  This reduces
the problem of calculating the persistence exponent to the calculation
of the decay rates \cite{Majumdar1}.

The  order  parameter  field  in  the OJK  theory  is  given  by  
$\phi  = {\rm{sgn}}(X)$. The autocorrelation function
\begin{equation}
A(T) = \langle\phi(0)\phi(T)\rangle 
= \langle{\rm sgn}X(0)\,{\rm sgn}X(T)\rangle,
\end{equation} 
for the field $\phi$ at a space point moving with the flow, 
is given by 
\begin{equation}
A(T)=\frac{2}{\pi}\sin^{-1}a(T),
\label{exat}
\end{equation} 
which follows from the fact that $\phi$ is a Gaussian field \cite{Oono}.
We will  determine the persistence probability $P_0(t)$ from  $A(T)$. 

We  briefly discuss the IIA \cite{Majumdar} and  use it  to obtain 
approximate values for the exponent $\theta$ following the development 
in ref.\,\cite{Sire}. In the scaling  limit, the interfaces occupy a 
very small  volume fraction  and as  a result  $\phi(T)$ takes  values 
$\pm 1$ almost everywhere. The correlator $A(T)$ can be written as
\begin{equation}
A(T)=\sum_{n=0}^{\infty}(-1)^n P_n(T),
\end{equation}
where $P_n(T)$ is the probability that the interval $T$ contains $n$ zeros
of $\phi(T)$. For $n\ge 1$, $P_n(T)$ is approximated by, assuming that the 
intervals between zeros of $X$ are independent, 
\begin{eqnarray}
P_n(t)&& = \langle T \rangle^{-1}\int_0^T dT_1\int_{T_1}^T
dT_2\cdot\cdot\cdot   \int_{T_{n-1}}^T  dT_n  \nonumber   \\  &&\times
Q(T_1)P(T_2-T_1)\cdot\cdot\cdot P(T_n-T_{n-1}) Q(T-T_n),\nonumber \\
\label{iiae}
\end{eqnarray}
where $\langle  T \rangle$  is the mean  interval size, $P(T)$  is the
distribution of  intervals between successive zeros and  $Q(T)$ is the
probability that  an interval of  size $T$ to  the right or left  of a
zero contains no  further zeros, i.e. $P(T)=-Q'(T)$ where the prime 
indicates a derivative. The  IIA has been made  in Eq.\ (\ref{iiae})  
by  writing the  joint distribution of zero-crossing intervals as  the 
product of the distribution of  single intervals. The Laplace  transform  
of Eq.\ (\ref{iiae}) leads to $\tilde{P}(s)=[2-F(s)]/F(s)$ where
\begin{equation}
F(s)=1+\frac{\langle T \rangle}{2}s[1-s\tilde{A}(s)]
\end{equation}
and $\tilde{A}(s)$ is the Laplace transform of $A(T)$.

It  is straightforward to  show that  the mean  interval size is 
$\langle T \rangle =  -2/A'(0)$. The expectation that  
$P_0(T)\sim e^{-\theta T}$ for  large   $T$  implies a  simple  
pole in  $\tilde{P}(s)$ at $s = -\theta$. The  persistence exponent 
$\theta$ is  therefore given by the first zero on the negative axis of 
the function
\begin{eqnarray}
F(s)=1-\frac{s}{A'(0)}\left[1-\frac{2s}{\pi}\int_0^\infty dT\,
\exp(-sT) \right. \nonumber \\
\times \left. \sin^{-1}a(T)\right].
\label{expo}
\end{eqnarray}

For further analysis is is useful to first extract the asymptotic 
behavior of the autocorrelation function $a(T)$ of the field $X(T)$. 
From Eqs.\ (\ref{2dres}) and (\ref{acfr}) we find, for $T \to \infty$, 
\begin{equation}
a(T) \sim  \cases{\exp(-T/2),\ \ \ \ \ \ \, d=2, \cr 
\exp(-5T/4),\ \ \ \ \ d=3.\cr}
\label{largeT}
\end{equation}  
We now turn to the results.

\section{Results}
The  term   $A'(0)$  can  easily   be  evaluated  to  give   $A'(0)  =
-\sqrt{17/2}/\pi$  in   $d=3$  and  $-\sqrt{2}/\pi$   in  $d=2$.  From
(\ref{expo})  $F(0)=1$,  and from  (\ref{largeT})  $F(s)$ diverges  to
$-\infty$  for   $s\rightarrow  -5/4$  and  $-1/2$  in   $d=3$  and  2
respectively.   Therefore, the  zero of  $F(s)$ lies  in  the interval
(-5/4,0)  and   (-1/2,0)  for  $d=3$  and   2  respectively.   Solving
(\ref{expo}) numerically for this zero,  we get the IIA values for the
persistence exponent as $\theta = 0.5034 \ldots$ for $d=3$ and $\theta
=  0.2406 \ldots$  in $d=2$.  In the  absence of  shear the  IIA gives
\cite{Sire}  $\theta =  0.2358\ldots$ in  $d=3$ and  $0.1862\ldots$ in
$d=2$, which agree quite well with simulations \cite{NewmanLoinaz}.

A very interesting feature of the $d=2$ result for $\theta$ is that it
is exactly  twice the value of  the exponent obtained  within the same
approximation  (i.e.\ using  OJK  theory  and the  IIA)  for the  {\em
unsheared}  problem   in  {\em  one}   space  dimension:  $\theta_{\rm
sh}^{d=2} = 2\theta_{\rm unsh}^{d=1}$. That  this must be so is easily
seen directly  from the form (\ref{2dfinal}) for  $a(t_1,t_2)$ for the
sheared  problem in  $d=2$. The  equivalent result  for  the unsheared
system   in   general   space   dimension  is   $a(t_1,t_2)   =   {\rm
sech}^{d/2}(T/2)$  \cite{Sire}. For  $d=1$ this  is identical  to Eq.\
(\ref{2dfinal}) apart  from an overall  factor 2 in  the (logarithmic)
timescale $T$.  It follows  that the relation $\theta_{\rm sh}^{d=2} =
2\theta_{\rm unsh}^{d=1}$ does  not require the IIA but  only that the
underlying  field $m$ (or,  equivalently, $X$)  be Gaussian,  i.e.\ it
requires use of the OJK theory  but not the IIA.  It is interesting to
speculate that  it might  even hold beyond  the OJK  approximation, in
which case one  might imagine that there is  a very simple explanation
for it. As yet, however, we have been unable to find one.

The autocorrelation function $A(t_1,t_2)$ is also interesting.  In the
limit  $t_2  \gg  t_1$   that  defines  the  autocorrelation  exponent
$\lambda$  \cite{Bray}, via  $A(t_1,t_2) \sim  (t_1/t_2)^\lambda$, the
quantity $a(t_1,t_2)$ is small and Eq.\ (\ref{exat}) can be linearised
in  $a(t_1,t_2)$ to  give,  from  Eq.\ (\ref{largeT})  with  $T =  \ln
(t_2/t_1)$,
\begin{equation}
A(t_1,t_2) \sim \cases{(t_1/t_2)^{5/4},
\ \ \ \ \ d=3, \cr
(t_1/t_2)^{1/2},\ \ \ \ \ d=2.}
\label{lambda}
\end{equation}  
These results  give $\lambda = 5/4$  for the sheared  system in $d=3$,
compared to $\lambda =3/4$ in the unsheared system \cite{Bray}, whereas
for $d=2$ the autocorrelation exponent takes the same value, $\lambda =
1/2$, in  both cases. We should  repeat the caveat that, for $d=2$, the
simple  power-law  form (\ref{lambda})  requires  the  limit $t_1  \to
\infty$, $t_2 \to \infty$ with  $t_2/t_1$ fixed but large. If $t_2 \to
\infty$  for  fixed  $t_1$,  Eq.(\ref{2dres}) gives  $a(t_1,t_2)  \sim
(t_1/t_2)^{1/2}  [\ln(\gamma t_2)/\ln(\gamma  t_1)]^{1/4}$, which does
not  have a  simple  scaling form  (it  is not  simply  a function  of
$t_1/t_2$).

\section{Conclusion}
We have  studied the  effect of shear flow on the persistence exponent  
$\theta$,  for  a   system  with  nonconserved  scalar  order parameter, 
using an approximate analytical approach based on the OJK theory and 
exploiting the ``independent interval approximation''. The persistence 
is defined in a frame locally moving with the flow. 

The exponent $\theta$  is nontrivial and is increased  by the presence
of shear. This  implies that the shear accelerates  the change of sign
of the  fluctuating field. In  dimension $d=2$ we find  the intriguing
result that $\theta$ has a value  equal to twice that of the unsheared
system in $d=1$, within  the 0JK theory.  The autocorrelation exponent
$\lambda$  increases  in  the  presence  of shear  for  $d=3$  but  is
unchanged by the shear in $d=2$.

For  nonconserved dynamics  in the  absence of  shear,  experiments on
liquid crystals  have been performed to measure  both $\theta$ \cite{Yurke}
and $\lambda$ \cite{Mason}.  There is also  a recent experiment on the
measurement of a two-time correlation function in order-disorder phase
transition  in  Cu$_3$Au \cite{Andrei}.  Liquid crystal experiments  
are a possible candidate  for testing our predictions in  a model  with 
nonconserved order  parameter, and numerical simulations may also prove 
useful. On the analytical front, the method of the correlator 
expansion \cite{Ehrhardt} might be used to obtain a more accurate result 
for $\theta$ in $d=3$ than can be obtained using the IIA.  

\begin{acknowledgments}
The work of NR was supported by a Commonwealth Fellowship -- UK.
\end{acknowledgments}

\end{document}